\documentclass[twocolumn,pre,showpacs,letterpaper]{revtex4}
\usepackage{graphicx,amssymb,amsmath}
\usepackage{xcolor}
\usepackage{pstricks}
\graphicspath{{figure_clean_PRE_SIMU/}}
\begin{document}

\title{The role of dissipation in flexural wave turbulence: from experimental spectrum to Kolmogorov-Zakharov spectrum.}
\author{Benjamin Miquel, Alexandros Alexakis}
\affiliation{Laboratoire de Physique Statistique, Ecole Normale Sup\'erieure, Universit\'e Pierre et Marie Curie, CNRS, 24 rue Lhomond, 75005 Paris, France.}
\author{Nicolas Mordant}
\email[]{nicolas.mordant@ujf-grenoble.fr}
\affiliation{Laboratoire des Ecoulements G\'eophysiques et Industriels, Universit\'e Grenoble Alpes, Domaine Universitaire, BP53, 38041 Grenoble, France}
\affiliation{Institut Universitaire de France}


\begin{abstract}
The Weak Turbulence Theory has been applied to waves in thin elastic plates obeying the F\"oppl-Von K\'arm\'an dynamical equations. 
Subsequent experiments have shown a strong discrepancy between the theoretical predictions and the  measurements. Both the dynamical equations and the Weak Turbulence Theory treatment require some restrictive hypotheses. Here a direct numerical simulation of the F\"oppl-Von K\'arm\'an equations is performed and reproduces qualitatively and quantitatively the experimental results when the experimentally measured damping rate of waves $\gamma_\mathbf{k}= a + bk^2$ is used. This confirms that the F\"oppl-Von K\'arm\'an equations are a valid theoretical framework to describe our experiments. When we progressively tune the dissipation so that to localize it at the smallest scales, we observe a gradual transition between the experimental spectrum and the Kolmogorov-Zakharov prediction. Thus it is shown dissipation has a major influence on the scaling properties stationary solutions of weakly non linear wave turbulence.
\end{abstract}

\maketitle

\section{Introduction}
\subsection{The spirit of Weak Turbulence Theory}
The Weak Turbulence Theory (WTT) is, in its simplest version, a statistical description of the evolution of large ensembles of weakly non-linear dispersive waves. The scope of WTT is very large: gravity and capillary surface waves, sound waves, Alfven waves, plasma waves, internal waves, nonlinear optics, Bose-Einstein condensates~\cite{Newell, Nazarenko, Zakharov}. The object of this article is the case of flexural waves in a thin elastic plate. When energy sources and sinks are clearly separated in spectral space, the system is expected to exhibit a Richardson-like energy cascade: energy is transferred conservatively among scales. This energy transfer generates a continuous spectrum, often referred to as the Kolmogorov-Zakharov (KZ) spectrum and whose analytical expression is derived by using a multi-scale method~\cite{Newell, Nazarenko, Zakharov}. 

Due to the possibility of spatially extended measurements, the case of waves in an elastic plate is a valuable candidate to test WTT in details. So far, the theoretical predictions obtained by D\"uring \emph {et al.}~\cite{During} remained elusive in experiments~\cite{Mordant,Boudaoud}. In this article we report the results of numerical simulations of such wave turbulence in thin elastic plates. The purpose of this work is twofold. First we simulate the Foppl-von K\'arm\'an dynamical equations with realistic physical parameters in order to allow for a direct comparison with experiments. In this way, we expect to confirm that the properties of wave turbulence observed in our experiments is indeed embedded in the framework of these equations and do not require additional physics. In a second step we use the versatility of the numerical simulation to gradually decrease the dissipation in order to localize it at the smallest scales. In this way we fulfill the requirements of the WTT and thus expect our results to be in agreement with the theoretical predictions of D\"uring {\emph et al.}~\cite{During}. This will point out the role of dissipation in the observed data.

\subsection{Thin plates dynamical equations}

In order to derive a simple equation for flexion waves in an elastic plate, several hypotheses are typically made.
We consider a thin elastic sheet whose thickness $h$ is small compared to its size $L_x$ and $L_y$ in the two other directions $x$ and $y$. The plate is supposed to be flat at rest so that a point in the plate is labelled with its 2D cartesian coordinates $\mathbf{r}=(x,y)$. The motion of the plate is characterized by the field of normal deformation denoted $\zeta (\mathbf{r},t)$: in-plane displacements and in-plane inertia are neglected. The derivation of the dynamical equation requires the displacement $\zeta$ to remain of the order of magnitude of the thickness of the plate $h$. Furthermore the slope should remain small: $\left| \nabla \zeta \right| << 1$. The strain in the plate is supposed to remain small as well so that the linear bulk elasticity relations between strain and deformation hold~\cite{Landau,Amabili,Audoly}.

These hypotheses lead to the dynamical F\"oppl-Von K\'arm\'an (FVK) equations:
\begin{eqnarray}
\partial _{tt} \zeta &=& -\frac{E h^2}{12\rho(1-\sigma^2)}\Delta^2\zeta + \frac{1}{\rho}\left\{\zeta,\chi \right\}\\
\Delta^2 \chi &=&- \frac{E}{2\rho}\left\{\zeta, \zeta \right\} \label{eq_FVK}
\end{eqnarray}
where the physical properties of the material are described by the following coefficients: Young's modulus $E$, Poisson's ratio $\sigma$, the density $\rho$. The brackets $\left\{\cdot,\cdot\right\}$ denote the bilinear differential operator 
\begin{equation}
\left\{\zeta,\chi\right\}=\partial_{xx}\zeta\partial_{yy}\chi + \partial_{yy}\zeta\partial_{xx}\chi -2 \partial_{xy}\zeta\partial_{xy} \chi\,.
\end{equation}
The linear part of the wave equation provides the dispersion relation for vanishingly small wave amplitudes that are only due to flexion: $\omega=ck^2$ with $c=\sqrt{\frac{E h^2}{12 \rho(1-\sigma^2)}}$. For finite amplitudes, the stretching due to the deformations is no longer negligible: the Gaussian curvature $\left\{\zeta,\zeta\right\}$ acts as a source term for the Airy stress function $\chi$, yielding a cubic nonlinear term $\left\{\zeta,\chi \right\} = O(\zeta^3)$. The Fourier transform of equation~(\ref{eq_FVK}) yields:
\begin{equation}
\partial_{tt}\tilde{\zeta}_\mathbf{k} = -\omega_k^2 \tilde{\zeta}_\mathbf{k} - \int d^{6}\mathbf{k}_{123} \delta_{123}(\mathbf{k}) V^{0}_{123}\tilde{\zeta}_\mathbf{k_1}\tilde{\zeta}_\mathbf{k_2}\tilde{\zeta}_\mathbf{k_3}\label{eq_fvk_fourier} \, .
\end{equation} 
In this equation, we use the following conventions for the direct and inverse Fourier transform $\tilde{\zeta}_\mathbf{k}=\int{d^2\mathbf{r} \zeta(\mathbf{r})\mathrm{e}^{-\mathrm{j}\mathbf{k\cdot r}}}$ and $\zeta(\mathbf{r})=(2\pi)^{-2}\int{d^2\mathbf{k} \tilde{\zeta}_\mathbf{k}(\mathbf{r})\mathrm{e}^{\mathrm{j}\mathbf{k\cdot r}}}$, we use the shorthand notation $ \delta_{123}(\mathbf{k}) = \delta(\mathbf{k-k_1-k_2-k_3})$, and the kernel yields 
\begin{equation}
V^{0}_{123}=\frac{E}{2\rho(2\pi)^4}\dfrac{\left|\mathbf{k\times k_1}\right|^2\left|\mathbf{k_2\times k_3}\right|^2}{\left|\mathbf{k_1-k}\right|^4}\, .
\end{equation}
The different Fourier modes are independent in the linear approximation but are coupled when the cubic term comes into play. Despite the apparent complexity of equation~(\ref{eq_fvk_fourier}) some statistical properties of this deformation field are obtained analytically by the Wave Turbulence Theory for small amplitudes, as described in the following section. 

\subsection{WTT formalism applied to the FVK equations}

Following D\"uring \emph{et al.}~\cite{During}, we present the hypotheses assumed by WTT to derive the stationary spectrum of solutions of FVK equations.
 
\subsubsection{Hypotheses}
Wave Turbulence Theory aims at describing homogeneous systems for weak non linearities for which energy exchanges occur only between resonant sets of wavetrains. For these resonances conditions to be easily fullfilled, the limit of an infinite system  $L\rightarrow \infty$ is considered in the derivation of the WTT equations. Physically, this hypothesis requires finite-size systems to have a high modal density: hence, the frequency difference between adjacent modes is overwhelmed by the nonlinear broadening. Forcing and dissipation are generally considered for out-of-equilibrium cases. WTT demands that forcing and dissipation are well separated in Fourier space: in the canonical configuration, forcing acts at large scales corresponding to wavevectors below a given limit $k_F$, whereas dissipation is effective at small scales above some $k_d >> k_F$ cutoff wavenumber. In this manner, a range of wavenumbers delimited by $k_F$ and $k_d$ (called the inertial range, or the transparency window) exists where both forcing and dissipation can be neglected and where energy is conservatively transferred. Hence, the energy flux $\phi$ through the scales is constant. The limit of small wave amplitude is required and has the following consequence:  energy exchanges are dominated by resonant waves that involve the smallest number of waves, i.e. 4-waves resonances in our case. In this way, the modulation of the wavefield induced by the nonlinear interactions is slow compared to the period of the waves. This scale separation allows for a multi scale analysis.

\subsubsection{Kolmogorov-Zakharov spectrum for plates}
The so-called Zakharov solution is the stationary spectrum that is built by an energy flux $\phi$ (which has the dimension of mass/time$^{3}$) that flows through the system in the out-of-equilibrium case. We denote $E^{(2D)}(\mathbf{k})=\frac{1}{L^2}\left\langle \left|\partial_t\tilde{\zeta}_\mathbf{k}\right| ^2\right\rangle$ the power spectrum density of the normal velocity $v=\partial_t{\zeta}$ statistically averaged over realisations and time. For this quantity which has the dimension of length$^4$/time$^2$, the Zakharov solution yields~\cite{During}: 
\begin{equation}
E_{KZ}^{(2D)}(\mathbf{k}) = C \phi^{1/3}\ln^{1/3} {(k^*/k)}
\end{equation}where the dimensional factor $C$ is expressed as a function of the plate properties $h$, $\sigma$, the dispersion relation coefficient $c$, and a pure number $C_0$:
\begin{equation}
C=C_0\frac{ch}{\rho^{1/3}(1-\sigma^2)^{2/3}}\, .
\end{equation}
Isotropy is assumed in the system, so that integrated over angles spectrum $E^{1D}(k)=\int{E^{(2D)}(\mathbf{k})k\mathrm{d}\theta }$ and frequency spectra  $E(\omega)$ are considered. These spectra have the dimension of length$^3$/time$^2$ and length$^2$/time, respectively, and read for the Zakharov solution: 
\begin{eqnarray}
E_{KZ}^{(1D)}(k) & = & 2 \pi C \phi^{1/3}k\ln^{1/3} {(k^*/k)}\label{eq_spectrak_WTT}\\
E_{KZ}(\omega) & = & \frac{\pi C}{c}\phi^{1/3} \ln^{1/3} {(\omega^*/\omega)}\, . \label{eq_spectra_WTT}
\end{eqnarray}
One usually looks for power law solutions for this spectrum, thus the presence of the logarithmic term is unusual. The following argument accounts for the presence of this correction: when looking for power law solutions $E_{KZ}^{(2D)}(\mathbf{k})\propto k^a$, one finds that the exponent $a$ is degenerated between the out-of-equilibrium case and the zero-flux equilibrium situation, with the common value $a=0$. This degeneracy is raised by introducing a logarithmic correction~\cite{During}. The cutoff wavenumber $k^*$ and frequency $\omega^*$ can be related to dissipative phenomenon and plays the role of an UV cut-off~\cite{Nazarenko}.

\subsection{Article overview}
As we recall in section~\ref{sec_stationnary_regime}, the experimental observations suggest that $E^{(1D)}_\mathrm{exp} \propto \phi^{1/2}k^{-0.2}$ which is steeper than the prediction (\ref{eq_spectrak_WTT}). The origin of this discrepancy remains unclear. In this article, we present some numerical integration of the FVK equations described in section~\ref{sec_tecnical_details} together with a description of the experiments. We present in section~\ref{sec_stationnary_regime} a comparison between our experiments and our numerical simulations performed with realistic parameters (dissipation, size, \emph{etc.}). We show in section~\ref{sec_exp2KZ} the spectra obtained by numerical simulation when dissipation is localized above some cutoff wavenumber, a situation consistent with the hypotheses of the theory.

\section{Description of experiments}

\label{sec_tecnical_details}
\subsection{Experimental configuration}
\begin{figure}
\centering
\includegraphics[width=6cm]{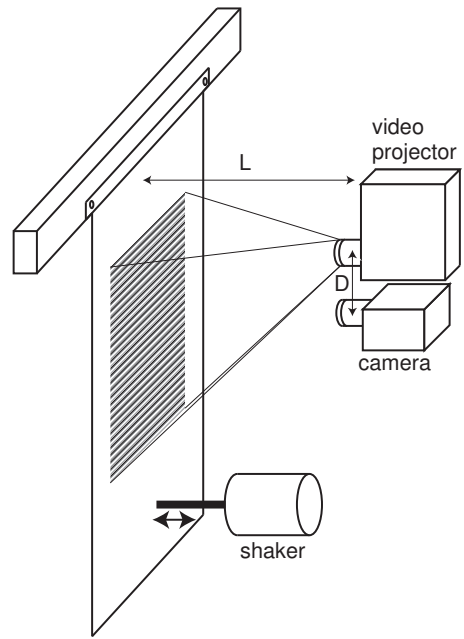}
\caption{Schematics of the experimental setup. A 0.4~mm thick, 2~m$\times$1~m stainless steel plate  is held vertical. Vibrations are excited by an electromagnetic shaker. Measurement of the deformation of the plate is achieved by projecting a grayscale pattern which deformation is recorded by a high speed camera~\cite{Cobelli1}.}
\label{setup}
\end{figure}

\paragraph{Setup: }The experimental setup is similar to the one used in~\cite{Miquel} (fig.~\ref{setup}). A stainless steel plate of dimensions ($2\mathrm{m}\times 1\mathrm{m} \times 0.4 \mathrm{mm}$) hangs under its own weight. The upper side is clamped and the three other edges move freely. An electromagnetic shaker fixed 40 cm above the bottom of the plate acts as a point source of waves. The deformation field is measured by using a profilometry technique developped by Cobelli et al.~\cite{Cobelli1}: a black and white sinusoidal grayscale pattern is projected on a large portion of the plate. The normal deformation of the plate yields some distortion of the pattern, that we record using an high-speed camera Photron SA1. The movie of the pattern distortion is demodulated into the movie of the displacement field. The frame rate is chosen between 5000 and 10000 frames per second depending of the magnitude of the forcing.

\paragraph{Modus Operandi:} After starting the vibration of the shaker, the study of the build-up transient regime reveals that the plate reaches a statistically stationary state after a delay of a few seconds. The deformation of the plate is then recorded during the stationary regime for a couple of seconds and this experiment is repeated so that ensemble averaging are computed during data processing. The decaying regime is recorded in a similar way by stopping the forcing after the stationary regime has been reached. The influence of forcing amplitude is investigated by tuning the forcing amplitude so that the ratio between strongest and weakest injected power is equal to 207; the forcing frequency is $f_0=30\ \mathrm{Hz}$.

\paragraph{Comparison with thin plates hypothesis:} The deformation of the plate driven by the shaker is of the same order of magnitude as its thickness for most of the wavelengths. However, some very large wavelength modes with amplitude greater than the thickness of the plate are observed. The large wavelength of these modes ensures yet that the slope remains small (rms value is of order of magnitude $0.05$). 

\paragraph{Comparison with WTT hypothesis:} Estimates of the damping time $T_d$~\cite{Miquel,arcasICA07,humbertEPL13} in plate reveal that dissipation exists at any scale and is not localized above some cutoff wavenumber, although there is no agreement on the analytic form of the damping coefficient. The experimental damping time for energy $T_d$ is measured in~\cite{Miquel} for wavenumbers ranging from $6\pi$ m$^{-1}$ to $60\pi$ m$^{-1}$. Over this range, $T_d$ is well approximated by a Lorentzian law~\cite{Miquel}: 
\begin{equation}
T_d(\mathbf{k})=(a+bk^2)^{-1}\label{eq_td}
\end{equation} 
with $a=0.73$ s$^{-1}$ and $b=6.3\times 10^{-4}$ m$^{2}$s$^{-1}$. Another difference with the WTT statements lies in the inhomogeneity of the force, that causes an inhomogeneity of the wavefield. The other features of the system are true to the spirit of wave turbulence. The frequency quantization due to the finite size of the system disappears as the modes are broadened by the nonlinearities so that the system behaves as an infinite system. Boundaries account for dissipationless reflexions. Finally, the double time-scale separation is preserved in our system: waves oscillate rapidly compared to their energy exchange, which is itself rapid compared to dissipation~\cite{Miquel3}.

\subsection{Numerical simulation of the F\"oppl-Von Karman equations}

\subsubsection{Equation and algorithm description} 

A forcing term $\mathcal{F}_\mathbf{k}(t)$ and a linear dissipative term $-\gamma_\mathbf{k} \partial_t \tilde{\zeta}_\mathbf{k}$ are added to eq.~(\ref{eq_fvk_fourier}) so that their impact on the shape of the spectrum are investigated: 
\begin{equation}
\partial_{tt}\tilde{\zeta}_\mathbf{k} = - c^2 k^4 \tilde{\zeta}_\mathbf{k}- \gamma_\mathbf{k} \partial_t \tilde{\zeta}_\mathbf{k}+\mathcal{F}_\mathbf{k}(t)  + \mathcal{N}_\mathbf{k}\left(\left\{\tilde\zeta\right\}\right)  \label{eq_fvk_fourier_damping_forcing} \, , \end{equation}
where $\mathcal{N}_\mathbf{k}$ denotes the nonlinear term of eq.~(\ref{eq_fvk_fourier}).
The integration of equation~(\ref{eq_fvk_fourier_damping_forcing}) is performed in a $T_2$ square (periodic in both directions) with an second order Runge-Kutta scheme and a pseudospectral method: the linear part is analytically propagated in Fourier space as the nonlinear part is evaluated in real space. The fields $\zeta$, $\partial_t \zeta$ and $\chi$ are dealiased with a 2/3 factor. The resolution is set between $192^2$ and $768^2$. The ratio between the largest and the smallest scales involved in the dynamics of a real plate does not exceed 100 for usual forcing amplitude, so that the dynamics of a real plate is reproduced faithfully with a resolution of $384^2$. As in experiments, the velocity and deformation fields are recorded to allow further data processing.

\subsubsection{Numerical parameters } 

The dispersion relation coefficient $c$ is set for all the simulations to a common value $c=0.64 \mathrm{m.s^{-1}}$ previously measured in experiments.

\paragraph{Forcing: } 
Each mode labelled by its wavevector $\mathbf{k}$ is forced resonantly at its linear eigenfrequency $\omega_\mathbf{k}$, with a phase $\phi_\mathbf{k}$ chosen randomly at the initial time and kept constant at later times. The magnitude of the forcing follows a Gaussian law peaked around some low but finite wavenumber $k_f=5\pi\textrm{m}^{-1}$ with a width $\sigma_k=2\pi\textrm{m}^{-1}$ or $4\pi\textrm{m}^{-1}$. The phase $\psi_{\mathbf k}$ of the mode is chosen randomly with the constraint $\psi_\mathbf{k} = \psi^*_\mathbf{-k}$. This yields:
\begin{equation}
\mathcal{F}_\mathbf{k}(t)= F_0 \mathrm{e}^{\mathrm{j}\psi_\mathbf{k}}\exp{\left(-\frac{\left( \left|\mathbf{k}\right|  - k_f\right)^2}{2\sigma_k^2}\right)} \cos\left(\omega_\mathbf{k} t + \phi_\mathbf{k} \right) \label{eq_forcing}
 \end{equation}

\begin{figure}[!htb]
\includegraphics[width=8.5cm]{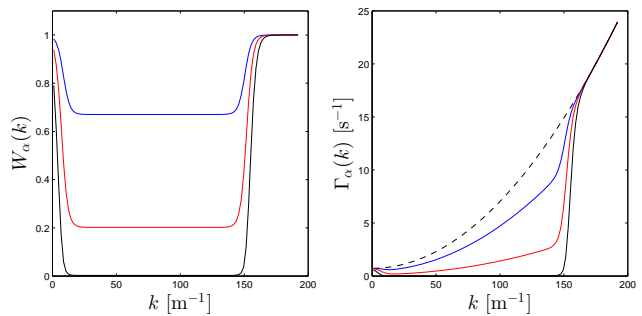}
\caption{ \label{fig_win_func} (Color online) Left: Band-rejecter function $W_{\alpha}(k)$ with $\alpha$ choosen equal to $0.1$, $0.4$ and $1.5$ for top to bottom solid lines. Cutoff wavenumbers are $k_l=9$ m$^{-1}$ and $k_h=150$ m$^{-1}$. Right: Corresponding tuned dissipation $\Gamma_{\alpha}(k)$ (solid lines) compared to experimental dissipation $\gamma_k^{EXP}$ (black dashed line) }
\end{figure}

\paragraph{Dissipation: } The role of dissipation is investigated by incorporating different damping rates $\gamma_\mathbf{k}$ in equation (\ref{eq_fvk_fourier_damping_forcing}). A first set of simulations $(\mathcal{S}_{EXP})$ uses realistic parameters to mimic the plate. Hence we use the damping rate measured in~\cite{Miquel}: 
\begin{equation}
\gamma_\mathbf{k}^{\mathrm{EXP}}=a+bk^2
\end{equation}
In the second set of simulations, the dissipation is gradually filtered out using a band-rejecter filter $W_\alpha(k)$:
\begin{equation}
\Gamma_{\alpha}(k)=W_{\alpha}(k)\gamma_k^{EXP}\label{eq_filtered_dissipation}
\end{equation}
The band rejecter $W_\alpha$ is characterized by its low and high cutoff wavenumbers $k_l$ and $k_h$, and the damping parameter $\alpha$ 
\begin{align}
W_{\alpha}(k)=&\exp\left( - \alpha \left[\tanh\left( \frac{k-k_l}{5}\right)+1\right]\right. \nonumber\\ &\times \left. \left[\tanh\left(\frac{-k+k_h}{5} \right)+1\right] \right) \label{eq_W}\end{align}
The band-rejecter behaviour of $W_\alpha$ is depicted on fig.~\ref{fig_win_func}: it takes values close to one for wavenumbers outside of the range delimited by $k_l$ and $k_h$ whereas inside this range the function decays smoothly to a plateau given by the rejection factor $\exp\left( - \alpha \right)$. Dissipation can be continuously removed in the inertial range $k_l< k<k_h$ by adjusting $\alpha$, as illustrated in fig.~\ref{fig_win_func}

 \section{Comparison between numerical and experimental results}
 \label{sec_stationnary_regime}
 
To demonstrate that FVK equations reproduce with a good agreement the experimental observations published in~\cite{Miquel,epjb,Miquel3}, we present in this section the results obtained by the set of simulations $(\mathcal{S}_{\mathrm{EXP}})$ that uses realistic damping rate $\gamma^{EXP}_\mathbf{k}$. 

 \subsection{Stationary regime}
 
 \begin{figure}[!htb]
\includegraphics[width=8.5cm]{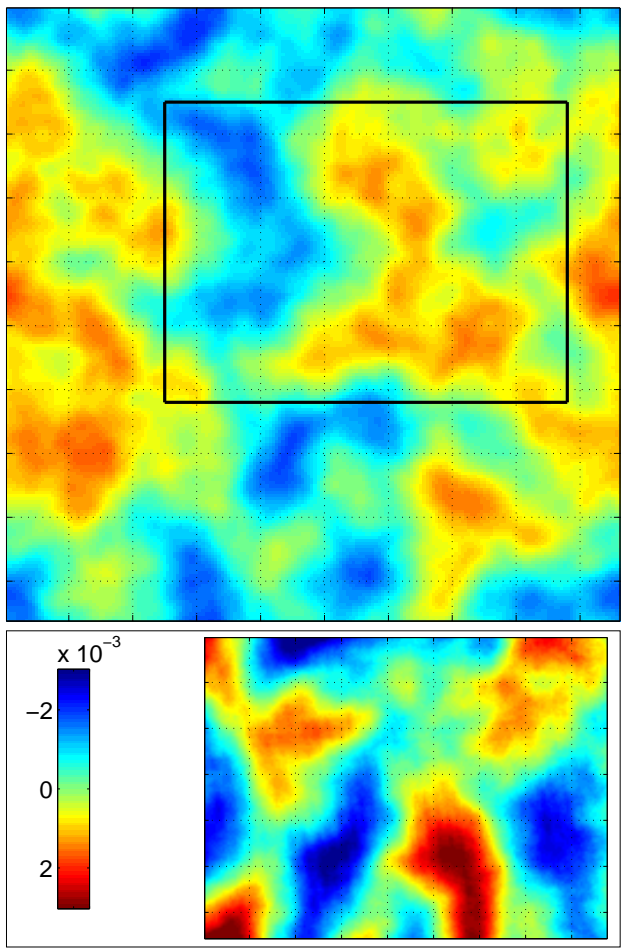}
\caption{ \label{fig_fields_sim_num} (Color online) Color coded snapshot of the deformation field $\zeta(\mathbf{r})$ (in meters). Upper figure: Numerical simulation. The thick black rectangle marks the size of the measurement window in the experimental setup. Bottom figure: the experimental deformation field (color plot) is embedded into a rectangle that marks the size of the plate (for the experimental data, the modes corresponding to the smallest $k$ that are not present in the simulation have been filtered out to ease the comparison). The forcing intensity is tuned in a manner that the numerical and the experimental cutoff wavenumber are comparable.}
\end{figure}

\begin{figure}[!htb]
\begin{pspicture}(8.5cm,6.5cm)
\includegraphics[width=8.5cm]{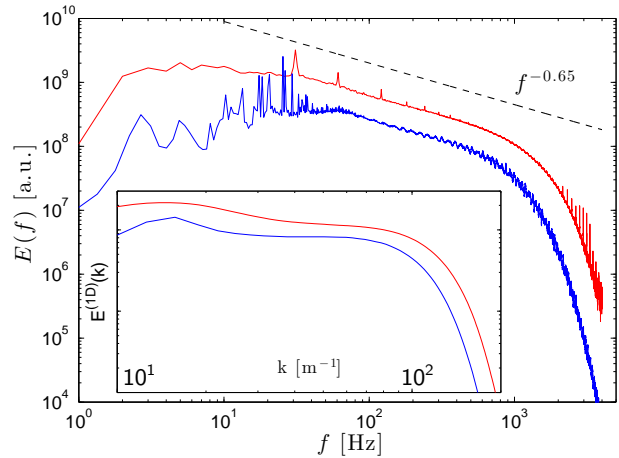}
\rput(-6.5,1.4){$\mathsf{10^1}$}
\rput(-2.8,1.4){$\mathsf{10^2}$}
\end{pspicture}
\caption{\label{fig_simu_manip} (Color online) Power spectrum density of the velocity field $E(f)$ (insert: power spectrum density in $\mathbf{k}-$space integrated over directions of $\mathbf k$: $E^{(1D)}(k)$). Top line: experimental spectra; bottom line: numerical spectra. Curves are vertically shifted for clarity.}
\end{figure}

A typical deformation field obtained by the simulation is displayed in figure~\ref{fig_fields_sim_num} and compared to a measured deformation field in a similar regime. Qualitatively, these two fields are very similar. The corresponding spectra displayed in figure~\ref{fig_simu_manip} show also a very good qualitative and quantitative agreement. 
They both exhibit a maximum corresponding to their respective forcing frequency. For intermediate frequencies, these spectra follow a power law $E(f)\propto f^{-0.6}$ and the spectra end in an exponentially decaying part $E(f)\propto \exp{(-f)}$. Note that the low-frequency part of the spectra is wider on the experimental data due to the possibility of large wavelengths to exist with the experimental boundary condition, whereas the fundamental frequency in the spatially periodic numerical domain is $1.0$ Hz. 
The corresponding spatial spectra $E^{(1D)}(k)$ are shown in inset of figure~\ref{fig_simu_manip}. They exhibit a similar behavior: a power law regime $E^{(1D)}(k)\propto k^{-0.2}$ follows a bump due to forcing and precedes an exponential decay $E^{(1D)}(k)\propto \exp{(-k^2)}$. The influence of injected power on spectra is studied by varying the amplitude of the forcing. 

\begin{figure}[!htb]
\includegraphics[width=8.5cm]{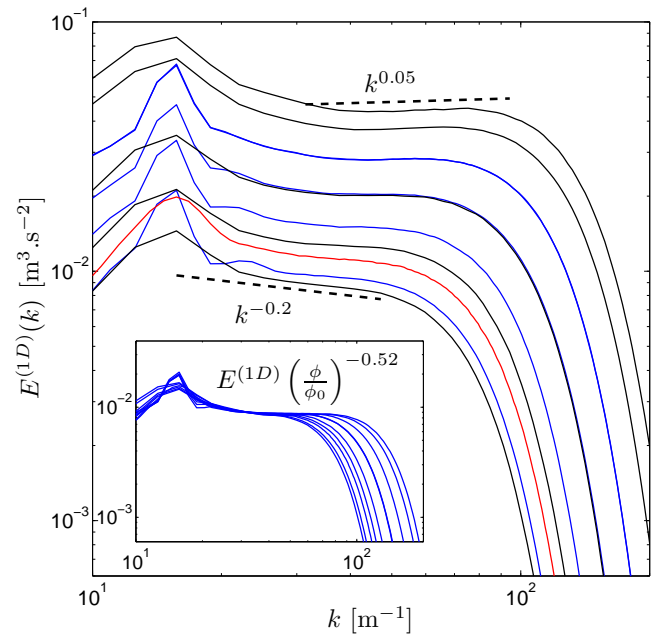}
\caption{\label{fig_scaling_simu} (Color online) Main figure: numerical spectra $E^{(1D)}(k)$ obtained with realistic parameters and for different forcing amplitudes. Different domain sizes are considered: $6\times6\mathrm{m}^2$ (red), $4\times4\mathrm{m}^2$ (blue), $2\times2\mathrm{m}^2$ (black). Bottom and upper dashed line are eyeguides for power laws $k^{-0.2}$ and $k^{0.05}$, respectively. The injected power per unit area is increased from bottom to top spectrum. Inset: the same spectra are rescaled with the injected power to the exponent $-0.52$.}
\end{figure}
We display in figure~\ref{fig_scaling_simu} a collection of spectra obtained with various forcing amplitudes and various plate sizes. As the injected power increases, spectra develop toward higher wavenumber and the scaling exponent varies slightly from $-0.2$ for weakest forcing to $0.05$ for strongest forcing. This slight variation could be due to the relatively small scale separation between forcing scales and dissipation scales. It results in a narrow scaling range whose exponent may be altered by the influence of forcing and/or dissipation.
Nevertheless the various spectra can be convincingly rescaled when dividing by $\phi^{0.52}$. This observation is at odds with the theoretical prediction of $\phi^{1/3}$. The experimental observations suggest an even larger exponent $\phi^{0.7}$~\cite{epjb}. Although not in perfect quantitative agreement, the numerical simulations and the experiments both show a similar disagreement with the WTT predictions.

\subsection{Space-time structure and nonlinear time}

\begin{figure}[!htb]
\includegraphics[width=8.5cm]{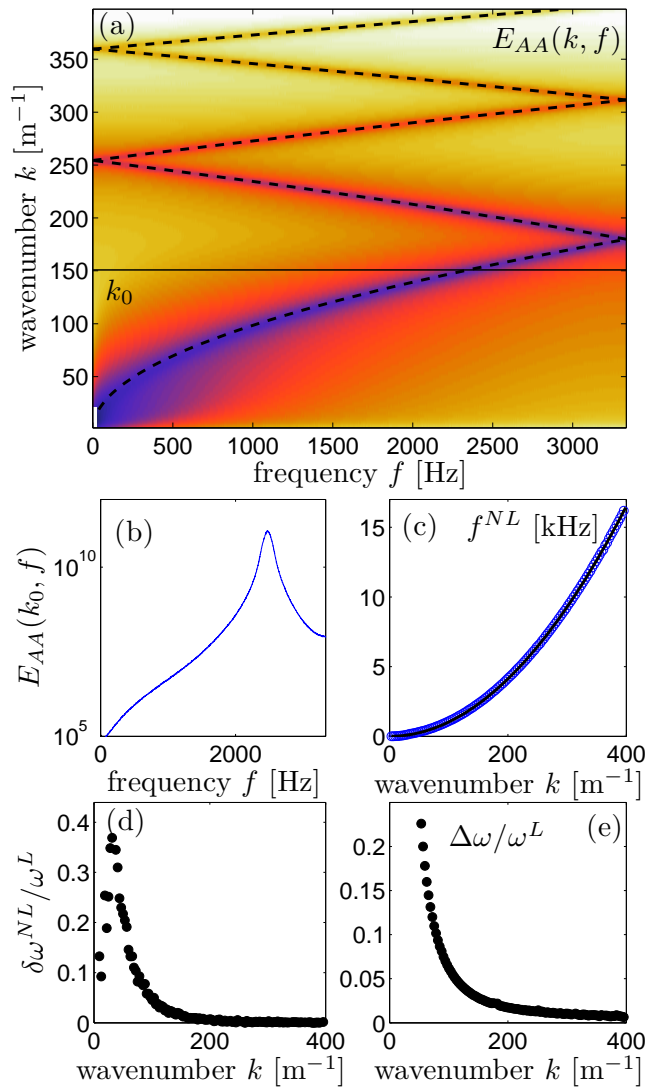}
\caption{\label{fig_simu_spectres_spa_temp} (Color online) (a): Angle-averaged time-space spectrum of the velocity $E_{AA}(k,f)$ plotted in a log color-chart as a function of frequency $f$ and wavenumber $k$. Although high frequencies are correctly resolved by the simulation, the field is not recorded at each time step but every $\delta t=0.15$~ms instead. It results in aliasing of the Fourier spectrum. Nevertheless the spectral information is still present in the picture. Dashed line: dispersion relation in the linear limit $f_L=\omega_L/2\pi  = ck^2/2\pi$ (purposely aliased). (b): cut in $E_{AA}(k,f)$ for wavenumber $k_0=150$m$^{-1}$ (materialized by the black horizontal line in (a)). This cut exhibits a maximum at the frequency $f(k_0)$. Figure (c) presents $f^{NL}_k=\omega^{NL}_k/(2\pi)$, the extracted crestline of $E_{AA}(k,f)$ as a function of $k$. The corresponding relative shift from the dispersion relation $\delta \omega^{NL} / \omega_L = \left( \omega(k) - \omega_L\right) /\omega_L$ is plotted in (d). (e) displays the width of the spectrum $\Delta \omega / \omega$ around the peak frequency. }
\end{figure}
In the linear limit, spatial modes of the deformation $\zeta_\mathbf{k}$ are not coupled. Unforced undamped modes oscillate with a constant amplitude $z_\mathbf{k}(0)$ at the pulsation obeying the dispersion relation:
\begin{equation}
\zeta_\mathbf{k}(t)=z_\mathbf{k}(0)\exp{(\mathrm{j}\omega_\mathbf{k} t)}
\end{equation}
WTT predicts a dual-role played by four-waves resonances: first, the pulsations of the modes $\omega^{NL}_\mathbf{k}$ are shifted from the dispersion in the linear limit $\omega_\mathbf{k}^{L}=ck^2$ with a factor $\delta \omega^{NL}_\mathbf{k}$ proportionnal to the energy of the waves:
\begin{equation}
 \omega^{NL}_\mathbf{k}= \omega^{L}_\mathbf{k}+\delta\omega^{NL}_\mathbf{k}\, .
\label{eq_shift_dispersion}
\end{equation}
Secondly, the modes are modulated in time by the slowly varying envelop $z_\mathbf{k}(t)$:
\begin{equation}
\zeta_\mathbf{k}(t)=z_\mathbf{k}(t)\exp{(\mathrm{j}\omega^{NL}_\mathbf{k} t)} \label{eq_modulation_lente}
\end{equation}
We define $T_{NL}$ as the characteristic time of the slow modulation $z_\mathbf{k}(t)$. The timescale separation hypothesis of WTT states that linear $T_L$ and nonlinear timescale $T_{NL}$ are strongly separated. We describe in the following two methods to verify this hypothesis. 

First, a qualitative conclusion may be drawn from the observation of the space-time spectrum $E^{(3D)}(\mathbf{k},\omega)$:
\begin{equation}
E^{(3D)}(\mathbf{k},\omega) =  \frac{1}{T}\frac{1}{L^2} \left\langle \left| \int_0^L d^2\mathbf{r}  \int_0^T{dt \zeta (\mathbf{r},t)    \mathrm{e}^{\mathrm{j}\omega t}} \mathrm{e}^{\mathrm{j}\mathbf{k}\cdot\mathbf{r}} \right|^2\right\rangle \end{equation}
Our isotropic forcing yields an isotropic spectrum as expected: $E^{(3D)}(\mathbf{k},\omega)$ does not depend on the direction of the wavevector $\mathbf{k}$ but only on the wavenumber $\left|\mathbf{k}\right|$ and the frequency $\omega$. Thanks to the isotropy, we can study the angle-averaged spectrum $E_{AA}$ (shown in fig.~\ref{fig_simu_spectres_spa_temp}):
\begin{equation}
E_{AA}(k,\omega)=\int{k E^{(3D)}(\mathbf{k},\omega) d\theta}\, .
\end{equation}

Figure (\ref{fig_simu_spectres_spa_temp}a) reveals that energy is located in the vicinity of the linear dispersion relation: for a given wavenumber $k$ (figure~(b)), the spectrum $E_{AA}(k,\omega)$ exhibits a maximum for $\omega=\omega^{NL}_k$ and a finite width $\Delta \omega$. The position of the crestline $\omega^{NL}_k$ is extracted by approximating $E_{AA}$ by a Gaussian function in the vicinity of the local maximum at given $k$:
\begin{equation} 
E_{AA}(k,\omega) \approx E_0(k) \exp\left[ -\frac{\left( \omega - \omega^{NL}_k\right)^2}{2(\Delta\omega)^2} \right] \,.\label{eq_sp_gaussien}
\end{equation}
The shifted dispersion relation $\omega^{NL}_k$ is shown  in fig. \ref{fig_simu_spectres_spa_temp}(c) together with the relative shift $\delta \omega^{NL}_k / \omega^L_k$ (d). Both the relative shift $\delta \omega^{NL}_k / \omega^L_k$ and the relative width $\Delta \omega / \omega_k$ (fig.~(\ref{fig_simu_spectres_spa_temp}e)) remain small in the cascade, supporting the validity of the time scale separation and of the weak non linearity. These observations are similar to experimental results reported in~\cite{epjb}.

An alternative yet equivalent way to define the nonlinear timescale $T_{NL}$ uses the autocorrelation function of the envelop: 
\begin{equation}
T_{NL}(\mathbf{k})=\int{  \left|\frac{\left\langle  \tilde{z}_\mathbf{k}(t)\tilde{z}^*_\mathbf{k}(t+\tau)\right\rangle}{\left\langle  \tilde{z }_\mathbf{k}(t)\tilde{z}^*_\mathbf{k}(t) \right\rangle} \right|\mathrm{d}\tau     } \label{eq_TNL}
\end{equation}
where the brackets $\left\langle  \cdot \right\rangle$ stand for statistical and time averaging. The absolute value is used to remove the fast oscillation at the frequency $\omega^{NL}_k$. The spectral widening $\Delta\omega$ and $T_{NL}$ are related by $\Delta\omega\propto1/{T_{NL}}$.

\begin{figure}
[!htb]
\includegraphics[width=8.5cm]{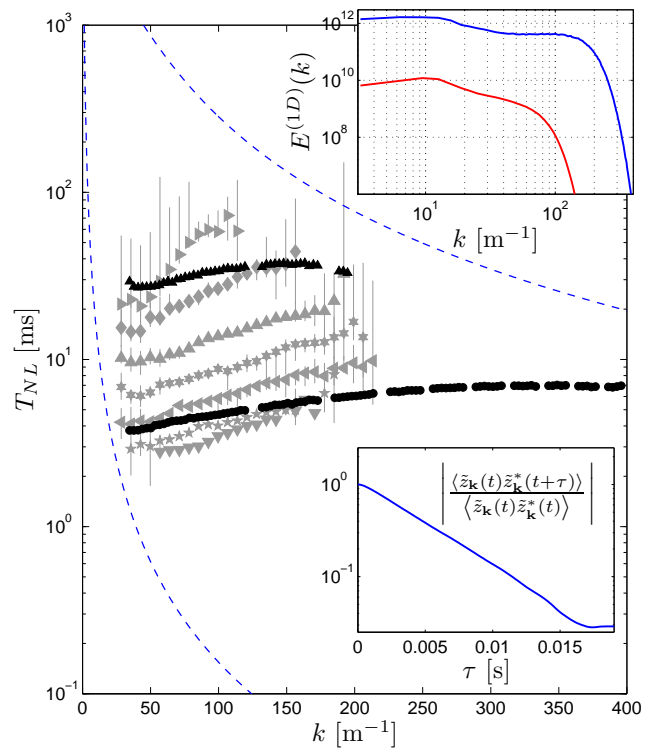}
\caption{(Color online) Comparison of the three time-scale: black plain symbols are the nonlinear time $T_{NL}$ obtained with equation (\ref{eq_TNL}) in numerical simulations (circles and triangle correspond to a strong and a weak forcing, respectively). The upper dashed line is dissipative timescale $T_d$ (see eq.\ref{eq_td}). The lower dashed line is the linear period of the waves $1/\omega_L(k)$. Experimental nonlinear times extracted from experimental data using a wavelet analysis described in \cite{Miquel3} are reproduced for comparison (grey symbols). Top inset: the spectrum $E^{(1D)}(k)$ is displayed to visualize the range of the various regimes for strong (blue or dark grey line) and weak (red or light grey) forcing. Bottom inset: exemple of autocoherence function of the slow modulation $\left|\frac{\left\langle  \tilde{z}_\mathbf{k}(t)\tilde{z}^*_\mathbf{k}(t+\tau)\right\rangle}{\left\langle  \tilde{z }_\mathbf{k}(t)\tilde{z}^*_\mathbf{k}(t) \right\rangle} \right|$ for $k=154$ m$^{-1}$.}
\label{fig_TNL}
\end{figure}
In the experiment, the dynamics is slightly different from the simulation by the fact that reflections occur at the plates boundary and only a portion of the plate is measured. Thus an indirect wavelet analysis of the motion described in~\cite{Miquel3} had to be used to compute $T_{NL}$ out of experimental data. Here, $T_{NL}$ is extracted in a more straightforward way by evaluating the autocorrelation as described in equation~(\ref{eq_TNL}). The three characteristic times of wave turbulence are plotted on figure~\ref{fig_TNL} together with experimental nonlinear timescales. First it can be seen that although dissipation occurs at all scales, it is weak so that a true scale separation occurs between dissipation timescale and the period of the waves. Second, the non linear time scale is also strongly separated both from dissipation and wave period. The experimental data points behave very closely to the numerical data showing again that the dynamics of experiments and simulation is similar and providing a validation of the wavelet method used for the experimental data in \cite{Miquel3}.

The very good signal over noise ratio in the case of the numerical simulation makes possible to measure $T_{NL}$  at small wavelengths in the dissipative region of the spectrum (exponentially decaying region). The spectrum of the corresponding runs is displayed in inset of figure~\ref{fig_TNL}. The fast exponential decay of the spectrum starts at $k$ close to 200. At this wavenumber the nonlinear time remains clearly separated from the dissipative time. It is somewhat surprising as this fast decay can very likely be attributed to dissipation. This early decay of the spectrum may also be attributed to an interplay with finite size effects. Nevertheless we cannot point out a direct influence of finite size so far.

\subsection{Decaying regime}

We study the decreasing stage of wave turbulence by considering a developed stationary spectrum as initial condition and by stopping the forcing. On a theoretical ground, a self-similarity argument exposed by Kolmakov in \cite{Kolmakov2} and applied to metallic plates in \cite{Miquel} yields an analytical expression for the dissipative region of the spectrum. In the absence of forcing, the variations in time of the spectrum are due to energy transfer among resonant waves and to dissipation. One looks for a self-similar solution $E^{(1D)}(k)=Ak_b^{\alpha}(t)g(k/k_b(t))$ where $k_b$ is a time dependent cutoff wavenumber, $A$ a constant that depends on the initial condition and $g$ a function of the dimensionless quantity $k/k_b$. The power $\alpha$ must be $3$ in order to match the homogeneity degrees in $k$ of the collision and dissipative terms. Dissipation dominates at the high wavenumber part of the spectrum and eventually yields $g(\xi) = \exp (-\xi^2)$. The time dependent spectrum is then expected to follow~\cite{Miquel}:
\begin{equation}
E^{(1D)}(k,t) = A k^3 \exp ({\color{red}-}k^2/k_b^2(t) )
\label{eq_Kolmakov}
\end{equation}
where an affine behaviour in time is predicted for $1/k_b^2(t)$:
\begin{equation}
k_b(t)^{-2}= Bt+C\, .
\label{eq_kb}
\end{equation}

\begin{figure}[!htb]
\includegraphics[width=8.5cm]{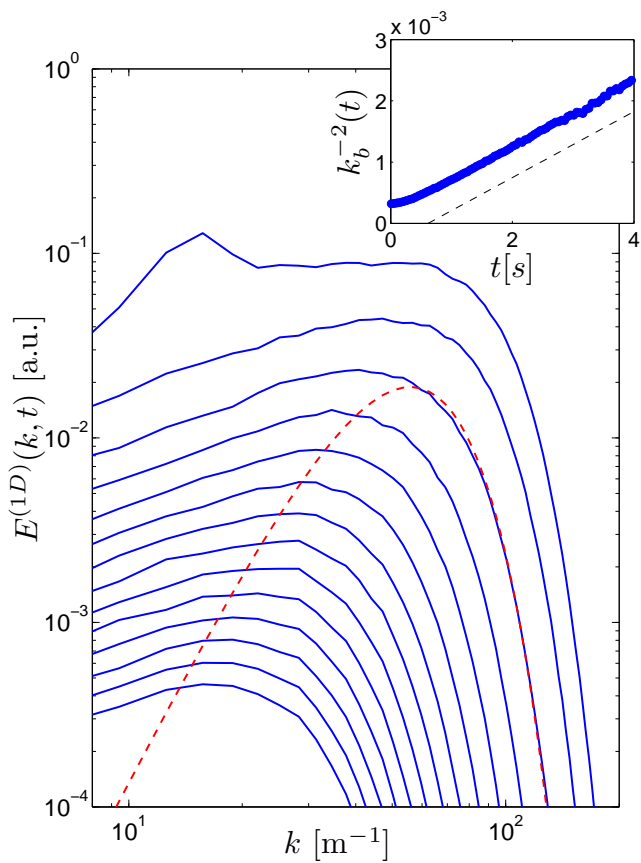}
\caption{(Color online) Blue plain lines : Integrated over angle power density spectra of velocity $E^{(1D)}(k)$ during decline stage. Upper curve: extinction of excitation $t=0$; curves are separated by $\Delta t= 0.3\ \mathrm{s}$ (time flowing downward). Red dashed line: fit for the dissipative part at $t=0.6\ \mathrm{s}$. Only the decaying part of the dashed line is expected to fit the data. Inset: dissipative cutoff $k_b^{-2}$ (blue dots) as a function of time compared to an affine law (dashed eyeguide)}
\label{fig_simu_declin} 
\end{figure}

Time-averaging is not possible to study this non-stationary regime. Instead, statistical averaging is performed as follow: a common forcing amplitude $F_0$ is chosen for the forcing of the form given in equation (\ref{eq_forcing}). Different phases $\phi_\mathbf{k}$ and $\psi_\mathbf{k}$ are chosen for each realization and the decaying stage is recorded. We display in fig.~\ref{fig_simu_declin} the averaged spectra for different delays after stopping the forcing. The dissipative parts of the spectra are fitted with equation (\ref{eq_Kolmakov}) and the extracted cutoff wavenumber obeys equation (\ref{eq_kb}) as depicted in the inset of figure~\ref{fig_simu_declin}. The behavior of the numerical simulation is again very similar to that of the experiment~\cite{Miquel}.

\subsection{Discussion}

We have presented above various statistical properties of wave turbulence simulated from the FVK equations using realistic parameters. We observe that all quantities are in very good quantitative agreement with the experimental observations. Hence we conclude that the FVK equations capture all the physics necessary to reproduce quantitatively the experiments. There is thus no need to invoke additional elements such as plate imperfections: for instance large-scale curvature due to a not perfectly flat plate (although commercial plates are indeed not perfectly flat) or residual anisotropic internal stress due to industrial fabrication (which is most likely present as no annealing of the plates has been performed). Such elements although present in real plates are not necessary to explain the discrepancy between observations (numerical or experimental) and theory. Two phenomena are likely candidates: dissipation and finite size. Simulations of plates with size 2, 4 or 6~m did not show any difference so that finite size effects can most likely be discarded in this precise case. To check the influence of dissipation we use in the following the versatility of the numerical simulations to decrease the dissipation rate in the inertial range so that to localize it at the smallest scales as is required for a strict application of the WTT.

\section{From experimental to KZ spectra}
\label{sec_exp2KZ}

\subsection{Energy Flux}
Equation~\ref{eq_FVK} formally conserves the mean energy per unit surface:
\begin{equation}
\mathcal{E}=\frac{h}{S}\int_S{d^2S \left[ \frac{1}{2}\rho c^2(\Delta\zeta)^2 + \frac{1}{2}\rho  v^2 + \frac{1}{2E}(\Delta \chi)^2         \right]       }\, .
\end{equation}
 This expression contain only quadratic terms in $\zeta$ and $\chi$ so that the total energy $\mathcal{E}$ is expressed as the sum of the energy $E_\mathbf{k}$ of the wavevectors $\mathbf{k}$:
\begin{eqnarray}
\mathcal{E}&=&  \sum_\mathbf{k} {E_\mathbf{k}} \\
    &=&  \sum_\mathbf{k} { \frac{1}{2}\rho c^2 k^4\left|\tilde{\zeta}_\mathbf{k}\right| ^2 + \frac{1}{2}\rho  \left|\tilde{v}_\mathbf{k}\right|^2 + \frac{1}{2E}k^4\left|\tilde{\chi}_\mathbf{k}\right|^2           } \, .
\end{eqnarray}
By restraining the summation to wavenumbers smaller a given $k$, we define the cumulative energy $\mathcal{E}_k=\sum_{\left|\mathbf{q}\right|< k} E_\mathbf{q}$ contained in the sphere of diameter $k$. The energy budget relates the variation of the energy contained in the sphere of radius $k$, the cumulative injected power $I_k$, the cumulative dissipated power $D_k$ and the outgoing energy flux $\Phi_k$. The injected and dissipated powers are explicitly computed using:
  \begin{eqnarray}
I_{k} &=& \left\langle\sum_{\left| \mathbf{q} \right| < k} {\mathcal{F}^*_\mathbf{q}(t)\tilde{v}_\mathbf{q}(t)}\right\rangle \label{eq_I}\\
D_{k} &=& \left\langle\sum_{\left| \mathbf{q} \right| < k} {\gamma_\mathbf{q} \left|\tilde{v}_\mathbf{q}\right|^2}\right\rangle\, , \label{eq_D}
\end{eqnarray}
where the brackets $\left\langle\cdot\right\rangle$ denote time averaging.
In the stationnary regime, the outgoing flux of energy balances the injected and dissipated energy:
\begin{equation}
\Phi_k =   I_k-D_k\, .  \label{eq_P}
\end{equation}

\begin{figure}[!htb]
\includegraphics[width=8.5cm]{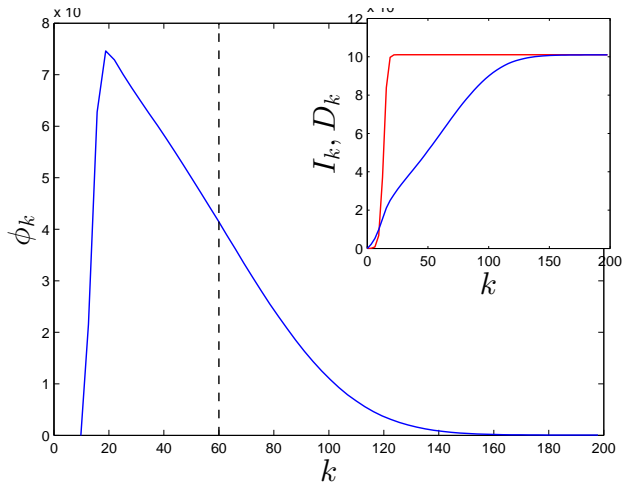}
\caption{\label{fig_flux_film3} (Color online) Inset: Injected (top red line) and dissipated (bottom blue line) power in the sphere of radius $k$ are plotted as functions of $k$. Main figure: energy flux $\phi_k$ as a function of $k$ (black plain line). Vertical dashed line indicates $k^*$ the cutoff wavenumber of the spectrum.
}
\end{figure}

This flux plays a central role in wave turbulence but it is extremely difficult to measure experimentally. Figure~\ref{fig_flux_film3} displays an example of an energy budget for a run taken from set $(\mathcal{S}_{EXP})$ (with realistic physical parameters).
The forcing is effective at low wave numbers only so that the cumulative injected power is constant with the radius of the sphere $q$.
By contrast, dissipation plays a significant role even for wavenumbers in the cascade, below the cutoff wavenumber $k^*$ that separates the power-law and the exponentially decaying regions of the spectrum. The flux $\Phi_k$ has decreased by a factor 2 when the cutoff is reached, violating the hypothesis of a constant flux through the cascade. This explains the observed disagreement of the measurements and simulation compared to the KZ spectrum.

\subsection{Transparency window}

\begin{figure}[!htb]
\includegraphics[width=8.5cm]{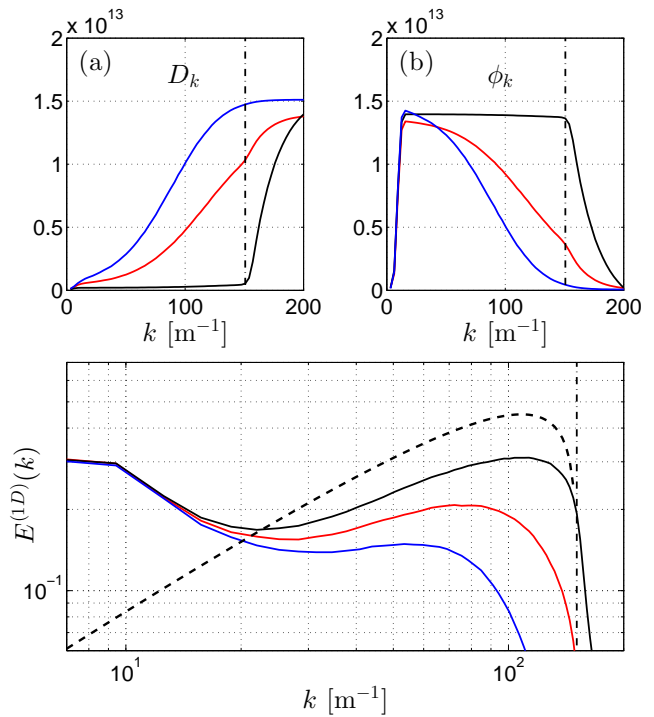}
\caption{(Color online)  (a): Dissipated power $D_k$ (see equation (\ref{eq_D})) as a function of the wavenumber $k$ using damping rate $\Gamma_\alpha(k)$ with $\alpha=0.1$ (top line), $0.4$ (middle line) and $1.5$ (bottom line). The dash-dotted vertical line indicates (as well as on the two following panels) the common high cutoff wavenumber parameter $k_h^*$ used in equation~(\ref{eq_W}) to define the filtered damping $\Gamma_\alpha(k)$. (b) corresponding Flux $\Phi_k$ over scale $k$ (see equation~(\ref{eq_P})) from bottom to top   $\alpha=0.1,0.4,1.5$. (c) power spectrum density $E^{(1D)}(k)$. The black thick dashed line indicates the Zakharov scaling $k\log(k_h/k)^{1/3}$ }
\label{fig_dissipation_filtree}
\end{figure}

The filtered dissipation $\Gamma_\alpha(\mathbf{k})$ defined in equation~\ref{eq_filtered_dissipation} is used so that dissipation is gradually suppressed in the inertial range. Figure~\ref{fig_dissipation_filtree} illustrates the runs obtained with $\alpha=0.1$, $0.4$ and $1.5$. As the damping rate (depicted in fig.~\ref{fig_win_func}) is decreased in the inertial range, dissipation becomes localized at high wavenumbers. Consequently the flux is seen to become constant through the cascade (fig.~\ref{fig_dissipation_filtree}a). The spectrum changes its shape so that to eventually resemble the Zakharov shape (fig.~\ref{fig_dissipation_filtree}b) (when setting $k^*=k_h$ in eq.~(\ref{eq_spectrak_WTT})). The correct wavenumber dependence of the spectrum $E^{(1D)}(k)$ is recovered when a genuine transparency window is created. This transition supports the intuitive idea that ``leaking" cascades lead to a spectrum steeper than the theoretical predictions of the conservative case~\cite{humbertEPL13,deikePRE14}.
\begin{figure}[!htb]
\includegraphics[width=8.5cm]{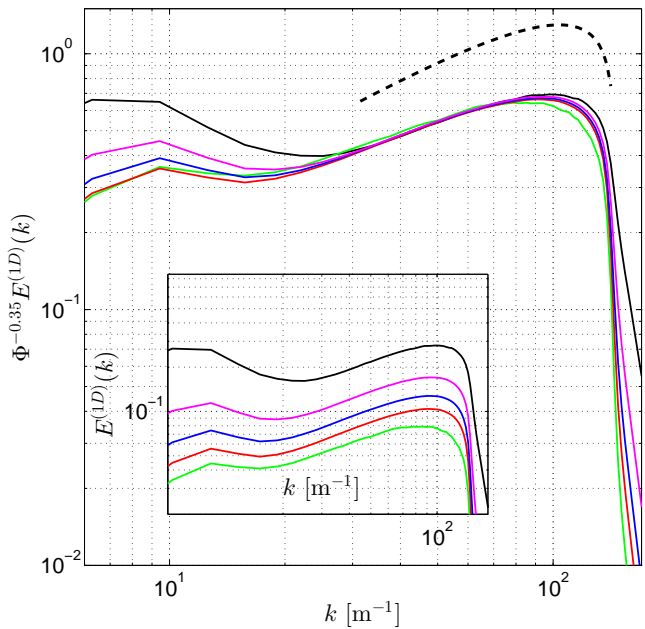}
\caption{ (Color online) Inset: spectra $E^{(1D)}(k)$ obtained for five different forcing amplitude and a transparent-like dissipation $\Gamma_{1.5}(k)$ are displayed by color plain lines. Main figure: the same spectra are rescaled with a power law of the flux $\Phi^{0.35}$ (color plain lines). Black dashed line indicates the Zakharov scaling $k\log(k_h/k)^{1/3}$  \label{fig_scaling_P}}
\end{figure}

In fig.~\ref{fig_scaling_P}, the dissipation is set to a transparent-like damping rate with $\alpha=1.5$ and the amplitude of the forcing is varied.  We investigate the scaling of the spectrum with the injected power. The various computed spectra collapse on a master curve when compensated by $\Phi^{0.35}$ which is very close to the Zakharov scaling $\Phi^{1/3}$. This is in strong contrast with the experimental dissipation for which a $\Phi^{1/2}$ is observed in numerics and $\Phi^{0.7}$ in experiments. It is remarkable that both the shape of the wave spectrum as a function of $k$ and the scaling with $\phi$ are restored when using a transparent dissipation. The scaling of the dissipation is seen to have a major impact on the wave spectrum. Not only the spectrum is steeper than the theory as a function of wavenumber but the scaling with the injected power is strongly altered as well, consistently with observations in gravity-capillary wave systems~\cite{deikePRE14}.

\section{conclusion}

The results of the numerical simulation with physical parameters relevant of real plates shows an extremely good agreement with the experimental observations. This supports the fact that the F\"oppl-von K\'arm\'an equations are an adequate theoretical framework to study the issue of wave turbulence in vibrated thin elastic plates. When the experimental dissipation is artificially tuned so that to tend to the theoretical configuration, one recovers the theoretical predictions both in the shape of the spectrum and the scaling with the injected power. Although the spectral variations of dissipation are seen to have a major impact on the wave spectrum in various systems~\cite{humbertEPL13,deikePRE14}, this influence is not taken into account so far in the Weak Turbulence Theory. Introducing realistic dissipation in the weak turbulence formalism is not straightforward as it has a direct influence on the mathematical handling of the resonances. Nevertheless empirical forms of the kinetic equations are used in operational tools of sea wave predictions incorporating various source and dissipation terms~\cite{model}. It suggests that despite the mathematical difficulty, introducing the empirical terms in the kinetic equation may be physically sound. The influence of such empirical terms on theoretical solutions is still a matter of investigations and it is seen here to have a strong influence both in terms of spectral shape and scaling with injected power.

\begin{acknowledgments}
This work was funded by the French Agence Nationale de la Recherche under grant TURBULON 12-BS04-0005. We thank Emmanuel Dormy for discussions about the numerical code.
\end{acknowledgments}

\bibliography{bibliov7}

\begin{thebibliography}{18}
\expandafter\ifx\csname natexlab\endcsname\relax\def\natexlab#1{#1}\fi
\expandafter\ifx\csname bibnamefont\endcsname\relax
  \def\bibnamefont#1{#1}\fi
\expandafter\ifx\csname bibfnamefont\endcsname\relax
  \def\bibfnamefont#1{#1}\fi
\expandafter\ifx\csname citenamefont\endcsname\relax
  \def\citenamefont#1{#1}\fi
\expandafter\ifx\csname url\endcsname\relax
  \def\url#1{\texttt{#1}}\fi
\expandafter\ifx\csname urlprefix\endcsname\relax\def\urlprefix{URL }\fi
\providecommand{\bibinfo}[2]{#2}
\providecommand{\eprint}[2][]{\url{#2}}

\bibitem[{\citenamefont{Newell and Rumpf}(2011)}]{Newell}
\bibinfo{author}{\bibfnamefont{A.~C.} \bibnamefont{Newell}} \bibnamefont{and}
  \bibinfo{author}{\bibfnamefont{B.}~\bibnamefont{Rumpf}},
  \bibinfo{journal}{Ann. Rev. Fluid Mech.} \textbf{\bibinfo{volume}{43}}
  (\bibinfo{year}{2011}).

\bibitem[{\citenamefont{Nazarenko}(2011)}]{Nazarenko}
\bibinfo{author}{\bibfnamefont{S.}~\bibnamefont{Nazarenko}},
  \emph{\bibinfo{title}{Wave Turbulence}} (\bibinfo{publisher}{Springer},
  \bibinfo{address}{Berlin}, \bibinfo{year}{2011}).

\bibitem[{\citenamefont{Zakharov et~al.}(1992)\citenamefont{Zakharov, LÕvov,
  and Falkovich}}]{Zakharov}
\bibinfo{author}{\bibfnamefont{V.~E.} \bibnamefont{Zakharov}},
  \bibinfo{author}{\bibfnamefont{V.~S.} \bibnamefont{LÕvov}}, \bibnamefont{and}
  \bibinfo{author}{\bibfnamefont{G.}~\bibnamefont{Falkovich}},
  \emph{\bibinfo{title}{Kolmogorov Spectra of Turbulence}}
  (\bibinfo{publisher}{Springer}, \bibinfo{address}{Berlin},
  \bibinfo{year}{1992}).

\bibitem[{\citenamefont{D\"uring et~al.}(2006)\citenamefont{D\"uring,
  Josserand, and Rica}}]{During}
\bibinfo{author}{\bibfnamefont{G.}~\bibnamefont{D\"uring}},
  \bibinfo{author}{\bibfnamefont{C.}~\bibnamefont{Josserand}},
  \bibnamefont{and} \bibinfo{author}{\bibfnamefont{S.}~\bibnamefont{Rica}},
  \bibinfo{journal}{Phys. Rev. Lett.} \textbf{\bibinfo{volume}{97}},
  \bibinfo{pages}{025503} (\bibinfo{year}{2006}).

\bibitem[{\citenamefont{Mordant}(2008)}]{Mordant}
\bibinfo{author}{\bibfnamefont{N.}~\bibnamefont{Mordant}},
  \bibinfo{journal}{Phys. Rev. Lett.} \textbf{\bibinfo{volume}{100}},
  \bibinfo{pages}{234505} (\bibinfo{year}{2008}).

\bibitem[{\citenamefont{Boudaoud et~al.}(2008)\citenamefont{Boudaoud, Cadot,
  Odille, and Touz\'e}}]{Boudaoud}
\bibinfo{author}{\bibfnamefont{A.}~\bibnamefont{Boudaoud}},
  \bibinfo{author}{\bibfnamefont{O.}~\bibnamefont{Cadot}},
  \bibinfo{author}{\bibfnamefont{B.}~\bibnamefont{Odille}}, \bibnamefont{and}
  \bibinfo{author}{\bibfnamefont{C.}~\bibnamefont{Touz\'e}},
  \bibinfo{journal}{Phys. Rev. Lett.} \textbf{\bibinfo{volume}{100}},
  \bibinfo{pages}{234504} (\bibinfo{year}{2008}).

\bibitem[{\citenamefont{Landau and Lifshitz}(1959)}]{Landau}
\bibinfo{author}{\bibfnamefont{L.}~\bibnamefont{Landau}} \bibnamefont{and}
  \bibinfo{author}{\bibfnamefont{E.}~\bibnamefont{Lifshitz}},
  \emph{\bibinfo{title}{Theory of Elasticity}} (\bibinfo{publisher}{Pergamon,
  New York}, \bibinfo{year}{1959}).

\bibitem[{\citenamefont{Amabili}(2008)}]{Amabili}
\bibinfo{author}{\bibfnamefont{M.}~\bibnamefont{Amabili}},
  \emph{\bibinfo{title}{Nonlinear Vibrations and Stability of Shells and
  Plates}} (\bibinfo{publisher}{Cambridge University Press},
  \bibinfo{year}{2008}).

\bibitem[{\citenamefont{Audoly and Pomeau}(2010)}]{Audoly}
\bibinfo{author}{\bibfnamefont{B.}~\bibnamefont{Audoly}} \bibnamefont{and}
  \bibinfo{author}{\bibfnamefont{Y.}~\bibnamefont{Pomeau}},
  \emph{\bibinfo{title}{Elasticity and Geometry: from hair curls to the non
  linear response of shells}} (\bibinfo{publisher}{Oxford University Press},
  \bibinfo{year}{2010}).

\bibitem[{\citenamefont{Miquel and Mordant}(2011{\natexlab{a}})}]{Miquel}
\bibinfo{author}{\bibfnamefont{B.}~\bibnamefont{Miquel}} \bibnamefont{and}
  \bibinfo{author}{\bibfnamefont{N.}~\bibnamefont{Mordant}},
  \bibinfo{journal}{Phys. Rev. Lett.} \textbf{\bibinfo{volume}{107}},
  \bibinfo{pages}{034501} (\bibinfo{year}{2011}{\natexlab{a}}).

\bibitem[{\citenamefont{Cobelli et~al.}(2009)\citenamefont{Cobelli, Maurel,
  Pagneux, and Petitjeans}}]{Cobelli1}
\bibinfo{author}{\bibfnamefont{P.~J.} \bibnamefont{Cobelli}},
  \bibinfo{author}{\bibfnamefont{A.}~\bibnamefont{Maurel}},
  \bibinfo{author}{\bibfnamefont{V.}~\bibnamefont{Pagneux}}, \bibnamefont{and}
  \bibinfo{author}{\bibfnamefont{P.}~\bibnamefont{Petitjeans}},
  \bibinfo{journal}{Exp. Fluids} \textbf{\bibinfo{volume}{46}},
  \bibinfo{pages}{1037} (\bibinfo{year}{2009}).

\bibitem[{\citenamefont{Arcas}(2007)}]{arcasICA07}
\bibinfo{author}{\bibfnamefont{K.}~\bibnamefont{Arcas}}, in
  \emph{\bibinfo{booktitle}{19th International Congress on Acoustics}}
  (\bibinfo{address}{Madrid}, \bibinfo{year}{2007}), Proceedings of the
  International Congresses on Acoustics.

\bibitem[{\citenamefont{Humbert et~al.}(2013)\citenamefont{Humbert, Cadot,
  D\"uring, Josserand, Rica, and Touz\'e}}]{humbertEPL13}
\bibinfo{author}{\bibfnamefont{T.}~\bibnamefont{Humbert}},
  \bibinfo{author}{\bibfnamefont{O.}~\bibnamefont{Cadot}},
  \bibinfo{author}{\bibfnamefont{G.}~\bibnamefont{D\"uring}},
  \bibinfo{author}{\bibfnamefont{C.}~\bibnamefont{Josserand}},
  \bibinfo{author}{\bibfnamefont{S.}~\bibnamefont{Rica}}, \bibnamefont{and}
  \bibinfo{author}{\bibfnamefont{C.}~\bibnamefont{Touz\'e}},
  \bibinfo{journal}{EuroPhys. Lett.} \textbf{\bibinfo{volume}{102}}
  (\bibinfo{year}{2013}).

\bibitem[{\citenamefont{Miquel and Mordant}(2011{\natexlab{b}})}]{Miquel3}
\bibinfo{author}{\bibfnamefont{B.}~\bibnamefont{Miquel}} \bibnamefont{and}
  \bibinfo{author}{\bibfnamefont{N.}~\bibnamefont{Mordant}},
  \bibinfo{journal}{Phys. Rev. E} \textbf{\bibinfo{volume}{84}},
  \bibinfo{pages}{066607} (\bibinfo{year}{2011}{\natexlab{b}}).

\bibitem[{\citenamefont{Mordant}(2010)}]{epjb}
\bibinfo{author}{\bibfnamefont{N.}~\bibnamefont{Mordant}},
  \bibinfo{journal}{Eur. Phys. J. B} \textbf{\bibinfo{volume}{76}},
  \bibinfo{pages}{537} (\bibinfo{year}{2010}).

\bibitem[{\citenamefont{Kolmakov}(2006)}]{Kolmakov2}
\bibinfo{author}{\bibfnamefont{G.~V.} \bibnamefont{Kolmakov}},
  \bibinfo{journal}{JETP Lett.} \textbf{\bibinfo{volume}{83}},
  \bibinfo{pages}{58} (\bibinfo{year}{2006}).

\bibitem[{\citenamefont{Deike et~al.}(2014)\citenamefont{Deike, Berhanu, and
  Falcon}}]{deikePRE14}
\bibinfo{author}{\bibfnamefont{L.}~\bibnamefont{Deike}},
  \bibinfo{author}{\bibfnamefont{M.}~\bibnamefont{Berhanu}}, \bibnamefont{and}
  \bibinfo{author}{\bibfnamefont{E.}~\bibnamefont{Falcon}},
  \bibinfo{journal}{Phys. Rev. E} \textbf{\bibinfo{volume}{89}},
  \bibinfo{pages}{023003} (\bibinfo{year}{2014}).

\bibitem[{\citenamefont{{WISE~group}}(2007)}]{model}
\bibinfo{author}{\bibnamefont{{WISE~group}}}, \bibinfo{journal}{Prog. Ocean.}
  \textbf{\bibinfo{volume}{75}}, \bibinfo{pages}{603} (\bibinfo{year}{2007}).

\end{thebibliography}

 \end{document}